 \documentclass[manuscript]{aastex}

\shorttitle{Bolometric corrections of SNe II\~-P}
\shortauthors{Bersten et al.}

\begin{document}

\title{Bolometric Light Curves for 33 Type II-Plateau Supernovae}

\author{Melina C. Bersten\altaffilmark{1} and Mario Hamuy\altaffilmark{1} }
\affil{Universidad de Chile, Departamento de Astronom\'{\i}a,
  Casilla 36-D, Santiago, Chile.}
\email{melina@das.uchile.cl}

\begin{abstract}

\noindent Using  data  of three well-observed type-II
plateau supernovae (SNe~II-P), SNe 1987A, 1999em  and 2003hn; and two
atmosphere  models by  Eastman et al. (1996) and Dessart $\&$ Hillier
(2005b) we derive calibrations for bolometric corrections and
effective temperature from BVI photometry. The typical scatter of
 the  bolometric correction is 0.11 mag. With these results we
obtain bolometric light curves and effective temperature evolution
for a sample of 33 SNe~II-P. The SN sample shows a range of 1 dex in
plateau luminosity and plateau durations from 75 to 120 days. 
Comparing the shape of the transition between the plateau and the 
radioactive tail, we find that the size of the drop is in the range 
of  0.8 to 1.12 dex.

\end{abstract}



\section{INTRODUCTION}

Nowadays we know that the  type II plateau supernovae, the most common type 
of supernovae (SNe) in nature \citep{2005A&A...433..807M, 2005A&A...430...83C,2009MNRAS.395.1409S}, are part of a larger group known as 
``core-collapse supernovae'' (CCSNe) ---which includes type Ib, 
type Ic, and other subclasses of type II SNe \citep{1997ARA&A..35..309F}--- sharing, in general terms, 
the same explosion mechanism \citep{2000Natur.403..727B,2007ApJ...656..372G}. 
Stars  which are born with  masses above  8 M$_\odot$  are  thought to end  
their lives as CCSNe \citep{2003ApJ...591..288H,2004MNRAS.353...87E}. The observational characteristics of these events 
strongly depend on the final stage of the progenitor object and on the properties of the 
circumstellar medium at the time of explosion. In the case of 
type II plateau supernovae (SNe~II-P) it is believed that  
the progenitors are red supergiants with thick hydrogen envelopes 
(generally several solar masses) \citep{1971Ap&SS..10...28G,1977ApJS...33..515F,1983Ap&SS..89...89L,2004Sci...303..499S}.
It is this particular feature and the fact that the star explodes in a very 
low-density environment \citep{2000ApJ...545..444B,2006ApJ...641.1029C}  which are responsible for the type II classification 
(H lines in the spectrum) and also for the plateau typing
(plateau-shaped light curve). 

The study of SNe II-P is critically important for understanding the range of progenitor 
masses, radii and energies which produce these objects.
One way to estimate such parameters is by comparing observations
(light curves, colors, spectra) with hydrodynamical models.
Some of the classical theoretical studies are those by
\cite{1983Ap&SS..89...89L} and  \cite{1985SvAL...11..145L}.
Attempts to compare these models with observations have been addressed by \cite{2003ApJ...582..905H}
and \cite{2003MNRAS.346...97N}.
However, these studies have not been satistactory owing to
1) the lack of good-quality data,
2) the usage of simplified relations between ill-defined and hard-to-measure photometric and spectroscopic
parameters, and
3) the fact that some of the models are based on simplified physical assumptions. 

In order to improve this situation, we have 1) enlarged the database of spectra
and light curves for 33 SNe II-P, and 2) developed our own hydrodynamical models using better
physics. Since our code produces bolometric light curves and effective temperatures,
it proves necessary to calculate these quantities from the observed photometry. 
Our final goal is to do this comparison for the 33 SNe,
but only three of them (SN~1987A, SN~1999em, and SN~2003hn) 
were observed over a sufficiently broad wavelength range
to allow the calculation of bolometric fluxes. 
The purpose of this paper is to 1) use the data for these three well-observed objects 
and explore the feasibility to derive a bolometric correction that could be 
applied for the remaining SNe with optical observations alone, and 2) derive
a calibration for effective temperature from optical colors.

Here we focus only on SNe~II-P since these objects possess extended, 
spherically symmetric hydrogen envelopes which smooth out
possible inhomogeneities arising from differences in the explosions themselves \citep{1989ApJ...341..867C,2001PASP..113..920L}.
Therefore, at least during the optically thick plateau phase, we expect a
photosphere radiating as a ``dilute'' blackbody whose properties are mainly 
driven by the photospheric temperature \citep{E96,D05}.
As the SN expands
the temperature drops monotically, so we expect a regular spectroscopic
evolution for this class of objects, and a well-behaved bolometric
correction with color during the plateau phase.
The inhomogeneities are expected to become more 
noticeable  by the end of the plateau, at which point the hydrogen
envelope is almost completely recombined, the photosphere lies near the 
center of the object, and the inner layers becomes visible. 

Usually the approach used to calculate bolometric light curves consists
in integrating the flux observed in all available photometric bands  
and making some assumption about the missing flux based on  spectroscopic 
observations or  models \citep{2003A&A...404.1077E,2006ApJ...641.1039F}. 
Here we propose a quantitative  method  to estimate 
the missing flux in the ultraviolet and infrared ranges using data from the three 
well-observed SNe, and two 
sets of atmosphere models. The derived bolometric corrections correlate
so well with optical colors that this calibration can be used to calculate 
bolometric light curves for many other SNe~II-P having $B V I$ photometry alone,
opening thus the possibility for a statistical analysis of the physical
properties of this type of objects.

We begin in \S ~\ref{sec:method} by describing the observational and theorical 
material used and  the procedure followed to calculate bolometric luminosities. 
In \S ~\ref{sec:results} we  derive calibrations 
for the  bolometric correction and effective temperature as a function of 
colors. Then in \S ~\ref{sec:sample} we apply these calibrations 
to obtain bolometric luminosities and effective temperatures for the 
sample of 33 SNe~II-P presented by \citet{H09}.
Finally in \S ~\ref{sec:conclusion} we outline our main results and conclusions
of this paper. In an upcoming paper \citep{BBH09} we will present theoretical light curves
from our hydrodynamical models in order to derive physical parameters for this set of SNe.

\section{Bolometric Flux for Calibrating Supernovae and Atmosphere Models } \label{sec:method}

In this section, we describe how we calculate bolometric fluxes for the three 
SNe~II-P which possess UV, optical and IR photometry, namely, SN 1987A, 
SN 1999em and SN 2003hn, and the two sets of atmosphere models available
in the literature \citep{E96,D05} (E96 and D05 hereafter, respectively).
This is our first step in order to examine if a bolometric correction  
can be derived from optical photometry alone. From now on, we will
refer to the SN evolution in terms of time or color indistinctly.
This is well justified during the plateau phase in which the SN 
atmosphere expands, cools and monotically turns reddder. 

\subsection{Observational and Theoretical Data} \label{sec:data}

In order to examine if a bolometric correction can be derived from 
optical colors, we made use of the three SNe~II that possess the
best wavelength and temporal coverage. Two of these are genuine SNe~II-P, 
SN 1999em and SN 2003hn, and the third is the famous SN 1987A which,
except for its peculiar light curve, shares most of the spectroscopic
properties of SNe~II-P. Most of the data of these three SNe was obtained at  
{\em Cerro Tololo Inter-American Observatory} (CTIO),  
{\em Las Campanas Observatory} (LCO), and the 
{\em European Southern Observatory} (ESO) at {\em La Silla}. 
For more details about the data reduction and instruments used see
\citet{H90} and \citet{Bouchet89} for SN 1987A, \citet{H09} for SN 1999em, 
and \citet{Kris08} for SN 2003hn. 
The photometric bands used in this analysis were $U B V R I Z J H K$ for 
SN 1999em and SN 2003hn, and $U B V R I Z J H K L M$ for SN 1987A. 

We adopted Cepheid distances for SN 1987A and SN 1999em with corresponding
values of 50 kpc \citep{2001ApJ...553...47F} and  11.7 Mpc \citep{Leo03}. For SN 2003hn we used a 
distance of 16.8 Mpc, as derived by Olivares et al. 2008 using the Standardized
Candle Method. We corrected the photometry for Galactic and host-galaxy 
extinction. We performed such corrections using Galactic visual absorptions 
of A$_V^{GAL}$=0.249 for SN 1987A, A$_V^{GAL}$=0.13 for SN 1999em, 
and A$_V^{GAL}$=0.043 for SN 2003hn \citep{SFD98}, assuming a standard
reddening law with R$_V$= 3.1 as given by Cardelli et al. (1989). 
The values used for host-galaxy absorption were, A$_V^{host}$=0.216 
for SN 1987A, A$_V^{host}$=0.18 for SN 1999em \citep{H01} and A$_V^{host}$=0.56
for SN 2003hn \citep{D08}, again assuming R$_V$= 3.1.

We also used in this analysis spectral energy distributions (SEDs) from two 
sets of SN atmosphere models (E96 and D05). These models depend on several 
parameters such as, luminosity, density structure, velocity and composition. 
For more details on the input parameters of such models the reader is
referred to \cite{E96} and \cite{D05}. We used a total of 61 model spectra from E96, 
and 107 from D05. We discarded 31 spectra from D05 which did not have 
enough UV coverage.

\subsection{Bolometric Luminosity Calculations} \label{sec:Lbol}

By definition, the bolometric luminosity is the integral of the flux 
over all frequencies. This integration can be done in a straight-forward way for 
the spectral models of E96 and D05 summing the flux over wavelength.
With the purpose to estimate bolometric corrections and colors for the
models we computed $B V I$  synthetic magnitudes using the filter 
transmission functions and zero points given by \citet{H01}.

For the three well-observed SNe the calculation of bolometric luminosities 
was performed from  reddening-corrected broadband magnitudes using the
values mentioned in section ~\ref{sec:data}. K-corrections were  negleected
due to the small redshifts involved. We began by computing a quasi-bolometric
light curve using all the available broadband data. The magnitudes were
converted to monochromatic fluxes 
at the specific effective wavelength of each filter using the
transmission  functions and zero points of the photometric system \citep{H01}.  
At epochs when a certain filter observation was not available,  
we interpolated its magnitude in time using the closest points. 
The total ``quasi-bolometric'' flux, $F_{qbol}$  was computed 
using  trapezium integration.

To estimate the missing flux in the UV and IR, $F_{UV}$ and $F_{IR}$, 
we fitted at each epoch a blackbody (BB) function to the monochromatic 
fluxes\footnote{These fits were restricted to the plateau phase where the envelope
of the SN was optically thick, and also to the transition to the nebular phase.
On the radioactive tail we did not use any UV or IR corrections.}. 
At early epochs the BB model provided very good fits to the fluxes in
all bands. As the photosphere became cooler the $U$-band flux started to
depart from the BB model in which cases we excluded this point from the fit.
At later epochs, subsequently the $B$-band and $V$-band data points showed 
the same behavior, departing from the BB model. The reason for this is 
related to the strong line blanketing that develops with time 
in that part of the spectrum. 

On the IR side the flux was extrapolated to $\lambda= \infty$ using the BB 
fits described above. The integral of that function between the longest 
observed effective wavelength and $\lambda= \infty$ was adopted as the
IR correction (F$_{IR}$). This correction increased with time but it always 
remained below  7$\%$ for the three SNe.
 
\begin{figure}
\epsscale{.80}
\plotone{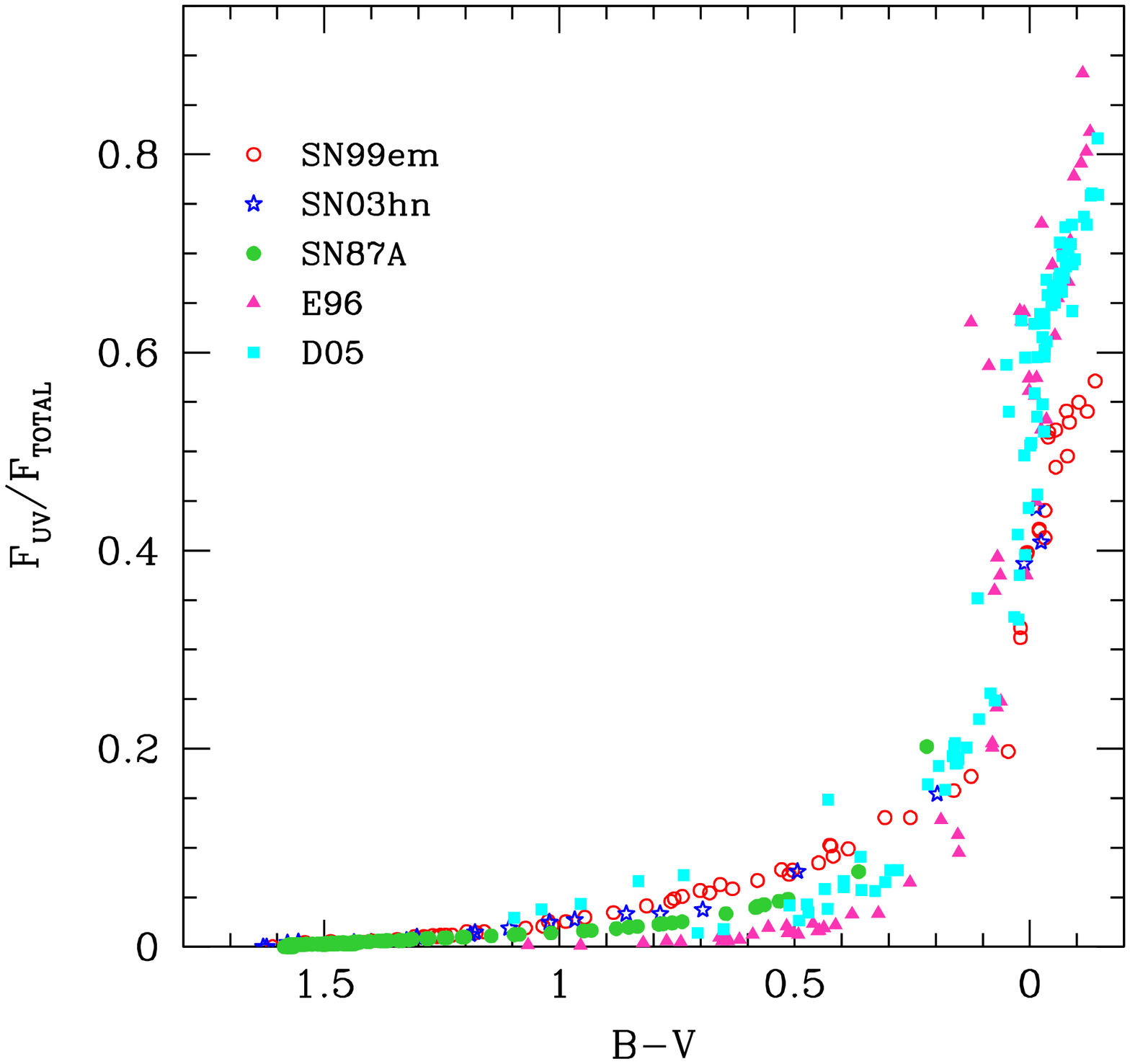}
\caption{UV contribution to the total flux as a function of
$(B-V)$ for the SN 1987A, SN 1999em and SN 2003hn and for 
the models of E96 and D05. At early times, 
when $(B-V) \lesssim 0.2 $, the UV flux represents a significant 
fraction of the total flux, implying larger uncertainties in the 
UV correction. \label{fig:fuv}}
\end{figure}

On the UV side, we extrapolated from the effective wavelength of the $U$ band
to $\lambda$=0 using the BB fit on all epochs except when the $U$-band flux 
fell below the BB model. In these cases, we extrapolated the $U$-band 
flux using a straight line to zero flux at 2000~\AA. Our choice of 
$\lambda$= 2000~\AA\ as the wavelength where the flux goes to zero
was based on the behavior of the atmospheric models for which the flux blueward  
of 2000~\AA\ is negligible in comparison with the total flux.
The integrated flux under the Plank function (or straight line) between the 
effective wavelength of the $U$ filter and $\lambda$=0 (or $\lambda$=2000 \AA) 
was taken as the UV correction (F$_{UV}$). The size of this correction 
relative to the total flux for the three SNe and the two sets of atmospheric 
models is shown in Figure ~\ref{fig:fuv} as a function of $(B-V)$.  The first
thing to note is the overall good agreement in F$_{UV}$ between the
observed SNe and the atmopshere models. Second, it is evident that
the UV correction is very important at early epochs and becomes nearly 
irrelevant at the latest epochs. Thirdly, note that in the very blue
end, where the UV corrections are of order 50-80\%, there is some
disagreement between the atmosphere models and SN 1999em. The spectral 
models suggest larger UV corrections and, as argued below, they are
more thrustworthy at these early epochs than the extrapolation of the 
broadband magnitudes. 
To prove this point, we calculated F$_{UV}$ for the models 
using the same technique that we employed for SN 1999em, i.e. by 
computing synthetic magnitudes for all passbands, converting them 
to monochromatic fluxes, and fitting a BB to the resulting points 
(instead of the direct integration of the SED).
In this case the UV correction proved closer to 
the UV correction derived from SN 1999em. We conclude that the UV
extrapolation using the BB fits to the broadband magnitudes 
at very early epochs underestimates somewhat F$_{UV}$, so we
ended up using only the atmosphere models at such epochs.
At later times ($B-V$$> -0.04 $), where the differences between data 
and theory become negligible, we adopted both the models and the observed data.

\begin{figure}
\epsscale{.80}
\plotone{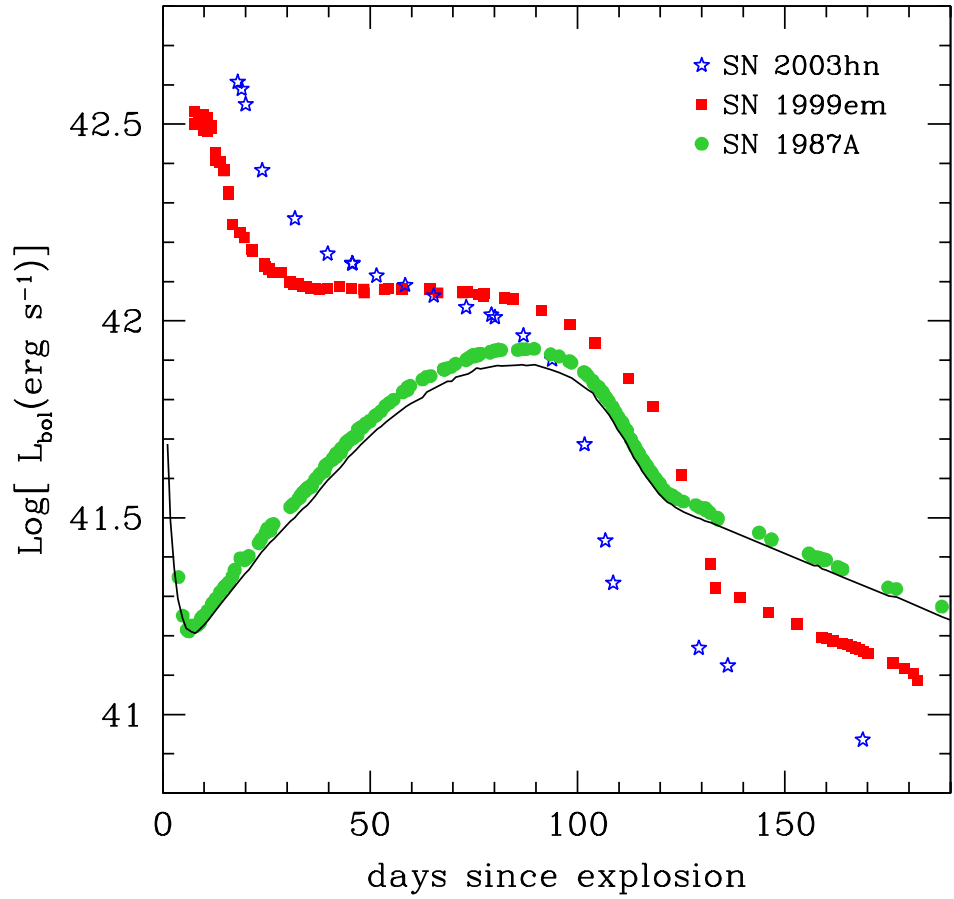}
\caption{Bolometric luminosity of SN 1987A, SN 1999em and SN 2003hn computed 
from the integration of broadband optical and near-infrared data plus  UV and IR 
contributions as explained in section~\ref{sec:Lbol}. For comparison we
included the bolometric luminosity of SN 1987A obtained  
by Suntzeff \& Bouchet (1990)(solid line).\label{fig:BLSNe}}
\end{figure}

The sum F$_{qbol}$+F$_{UV}$+F$_{IR}$ yielded the bolometic flux F$_{bol}$. 
Then we transformed flux into luminosity using the distances given in section 
\ref{sec:data}. The resulting bolometric luminosities for SN 1987A, SN 1999em 
and SN 2003hn are shown in Figure ~\ref{fig:BLSNe}. As a comparison, the solid line shows
the bolometric luminosity of the SN 1987A obtained by Suntzeff \& Bouchet
(1990). We found a very good qualitative agreement between both bolometric light curves for SN 1987A. 
There is a systematic difference which remains smaller than 0.04 dex at all times between both calculations. Such differences are consistent with the uncertainties estimated  for the use of  different datasets and different integration and interpolation scheme \citep{SB90}.

As it can be seen in Figure~\ref{fig:BLSNe}, the morphologies of the bolometric light curves for the three SNe are differents, specially that of SN 1987A which shows a broad  maximum, not observed in the classical SN II-P, and a less luminous light curve (up to the transition to the radioactive tail). It is  a well known fact that SN 1987A showed a peculiar light curve and this was because its progenitor was a compact blue supergiant that lead to a dim initial plateau and  to a light curve promptly powered by radioactivity \citep{1988ApJ...330..218W,1990ApJ...360..242S} . However, the three SNe showed a similar initial phase where the supernova rapidly faded and cooled until the outermost parts of the ejecta reached the temperature of hydrogen recombination (adiabatic cooling phase). A second phase can be distinguished for SN 1999em and SN 2003hn which corresponds to the plateau  where the brightness of the supernovae remained nearly constant  while hydrogen is recombining \citep{1971Ap&SS..10...28G}. The duration  and the slope of the light curve during this phase were different for each supernova. This is related to the properties of the progenitor object, mainly to the mass and radius of the hydrogen envelope. The shape of the light curve for  the SN 1987A during this phase was very different as mentioned above. It showed  a broad maximum characterize by a slow rise of $\sim$ 90 days followed by a more rapid decline for about 30 days. Finally, we can distinguish a third phase the radioactive tail which has a similar shape for the three SNe. Here, the luminosity has a  lineal behavior  and it is dominated by radioactive decay of $^{56}$Co. The luminosity in this part of the light curve is  a direct indicator of the mass of $^{56}$Ni synthesized in the explosion \citep{1989ApJ...346..395W}, in the sense of that more luminosity implies more $^{56}$Ni mass. We deduce from this that SN 1987A produced more  $^{56}$Ni  than the other two SNe.

\section{Calibrations}\label{sec:results}

\subsection{Bolometric corrections versus Color} \label{subsec:BC}

There are many SNe that lack IR and UV observations for which it is not 
possible to calculate the bolometric luminosity using 
the method described in the previous section. For these cases it
is necessary to know the bolometric correction required to convert
a $V$-band magnitude into a bolometric flux, i.e.,

\begin{equation}
BC = m_{bol} - [V - A_V],
\label{eq1}
\end{equation}

\noindent
where $A_V$ is the total visual extinction and $m_{bol}$ is the bolometric
magnitude in the Vega system. Note that, since BC is defined as a
magnitude difference, it is independent of the distance assumed for each 
object.

We calculated BC at all epochs for each of the calibrating SNe 
and for all of the SN models using the bolometric luminosities computed in
section \ref{sec:Lbol}. The bolometric fluxes were converted into Vega magnitudes 
in the following manner,

\begin{equation}
m_{bol}=  -2.5 \,  \,log_{10}\, \, F_{bol}\, +  11.64,
\label{eq2}
\end{equation}

\noindent where the zeropoint is obtained by integrating the SED
of Vega given by Hamuy(2001) and forcing the resulting magnitude
to vanish, i.e., m$_{bol}$(Vega)= 0.

\begin{figure}
\epsscale{.80}
\plotone{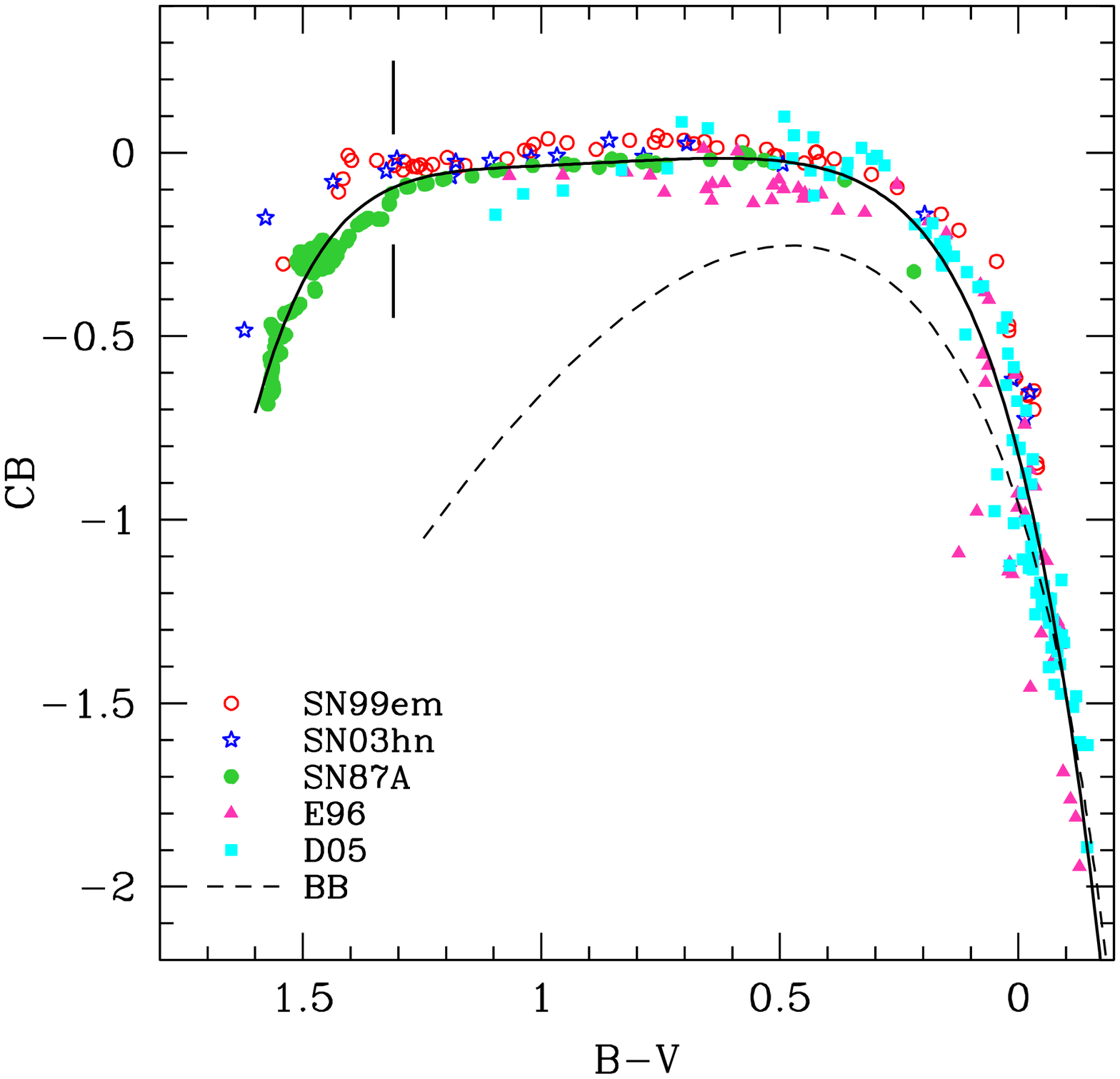}
\caption{Bolometric corrections versus $(B-V)$ for  SN 1999em (open circles), 
SN 2003hn (stars), SN 1987A (filled circles)  and the models of E96
 (triangles) and D05 (squares). The vertical lines indicate 
the color end at the the plateau phase.  The solid line  shows  a 
polynomial fits to the points.  The dashed curve correspond to the 
bolometric corrections of a blackbody spectrum.\label{fig:BC1}}
\end{figure}

\begin{figure}
\epsscale{.80}
\plotone{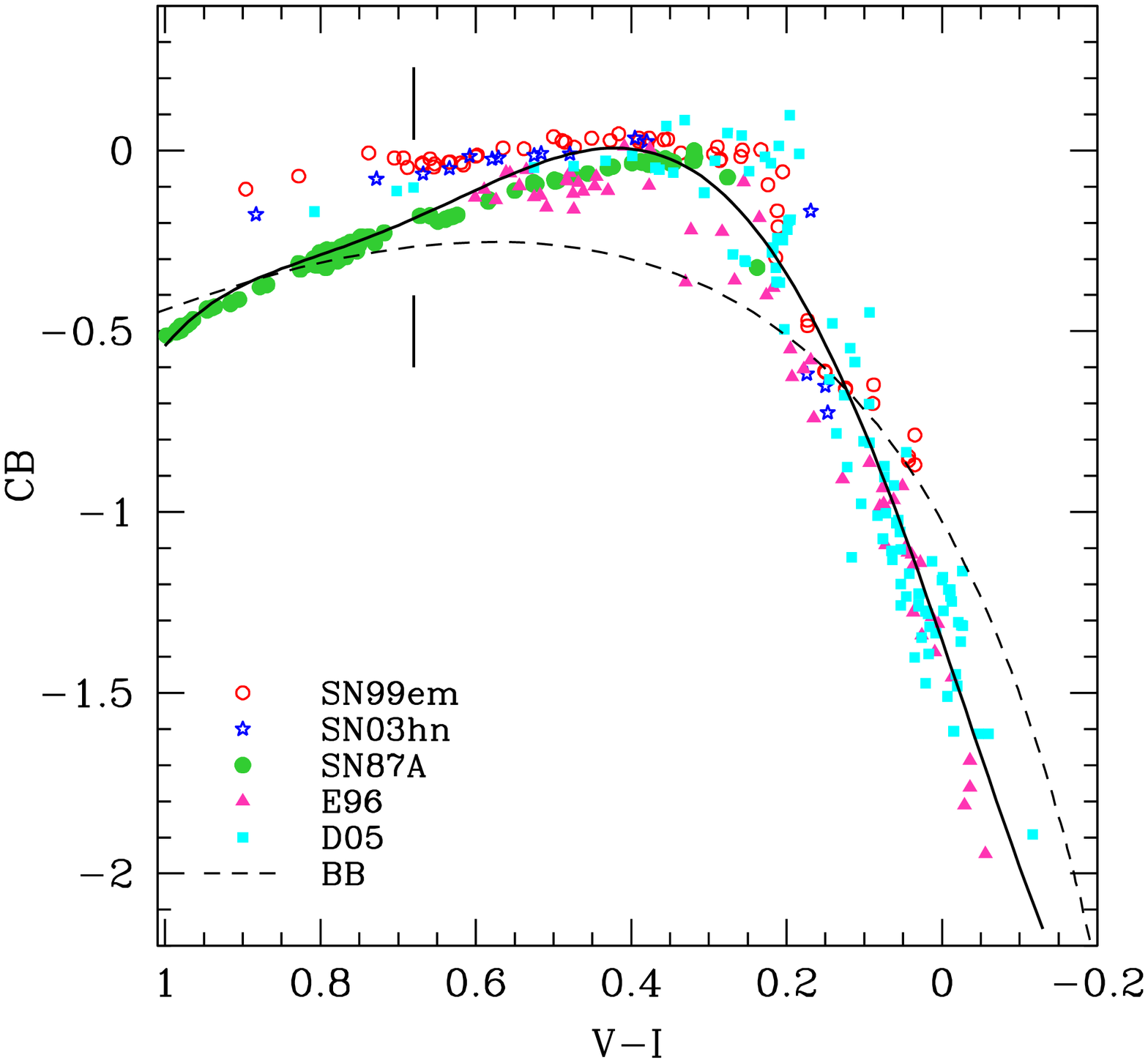}
\caption{Bolometric corrections versus $(V-I)$ for  SN 1999em (open circles), 
SN 2003hn (stars), SN 1987A (filled circles)  and the models of E96
 (triangles) and D05 (squares). The vertical lines indicate 
the color end at the the plateau phase.  The solid line  shows  a 
polynomial fits to the points.  The dashed curve correspond to the 
bolometric corrections of a blackbody spectrum. \label{fig:BC2}}
\end{figure}

\begin{figure}
\epsscale{.80}
\plotone{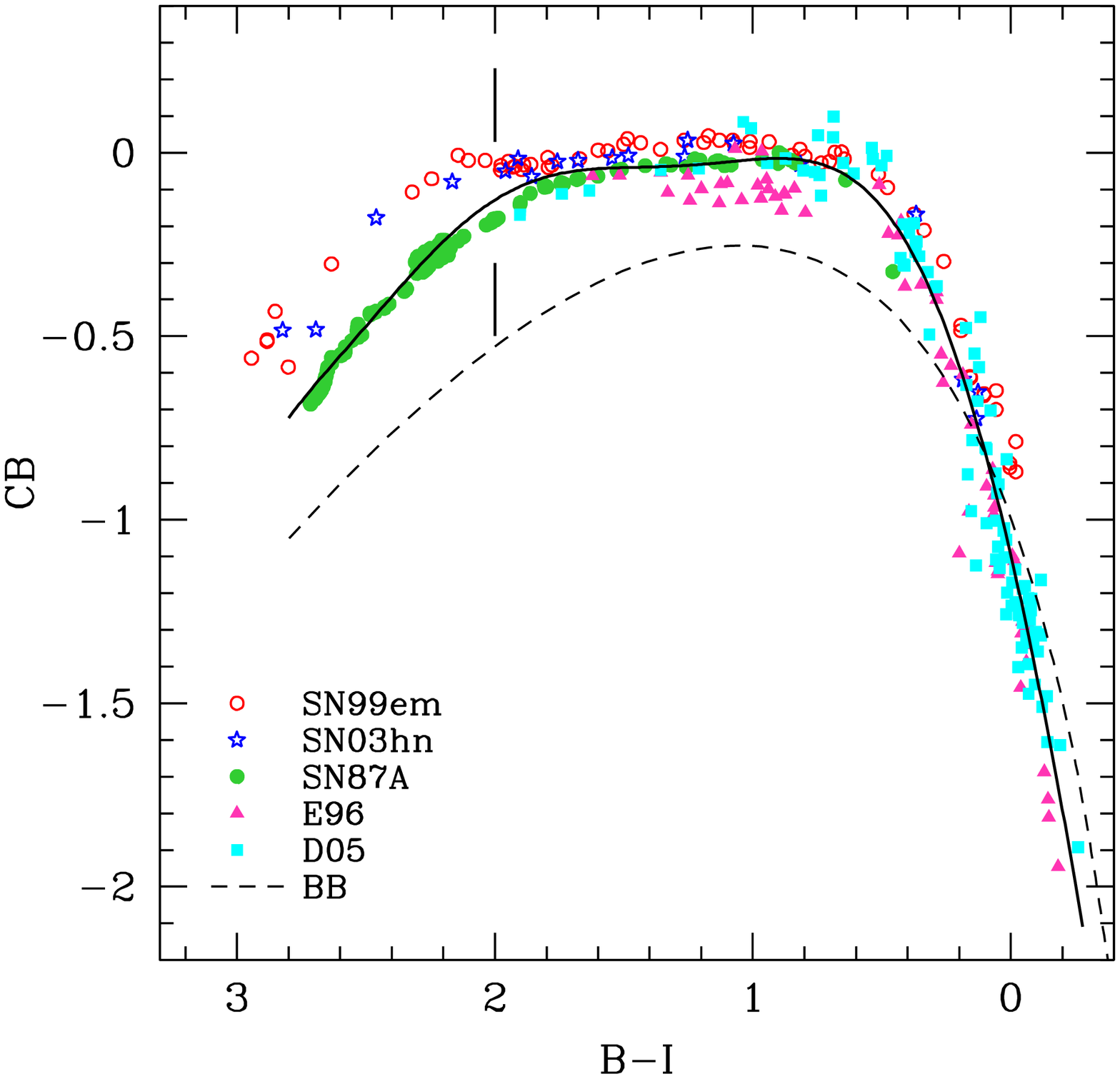}
\caption{Bolometric corrections versus $(B-I)$ for  SN 1999em (open circles), 
SN 2003hn (stars), SN 1987A (filled circles)  and the models of E96
 (triangles) and D05 (squares). The vertical lines indicate 
the color end at the the plateau phase.  The solid line  shows  a 
polynomial fits to the points.  The dashed curve correspond to the 
bolometric corrections of a blackbody spectrum. \label{fig:BC3}}
\end{figure}

We  analyzed the dependence of the BC on color, using
$(B-V)$, $(V-I)$ and $(B-I)$. The main reason why we did not try 
colors involving the $R$ band is because our sample \citep{H09}
has few SNe with $R$-band observations.
Figures \ref{fig:BC1},  \ref{fig:BC2} and  \ref{fig:BC3} show the resulting corrections 
as a function of the mentioned colors (corrected for dust) 
for SN 1987A, SN 1999em, SN 2003hn, and the models of E96 and D05. 
In each of these plots 
the vertical bar indicates the approximate color corresponding to the end 
of the plateau (see section \ref{sec:sample} for a precise definition of plateau phase).
 To the red of this mark are shown the BC  corresponding to the 
 transition between the plateau and the radioactive tail for the three SNe 
(no atmosphere models cover this phase). We do not include any of the nebular data 
in these diagrams, since the BB fits are not appropriate to extrapolate
UV or IR fluxes at these epochs.

These plots reveal a remarkable correlation between BC and intrinsic color, 
both for the objects and the models. It is very satisfactory that, even 
though SN 1987A has a very different light curve compared with normal SNe~II-P,
it matches quite well the behavior of the other two SNe and the models.
At very early times the BCs are quite large owing to the relatively
larger flux contribution of the UV. During most of the plateau 
the BC remains very small around a value of zero, which implies 
that the $V$ magnitude provides a very close proxy for the bolometric magnitude.
During the transition from the plateau to the radioactive tail (redward from 
the vertical bar) the BC starts to depart from zero due
to the larger flux contribution in the IR. At these phase
SN 1987A shows some discrepancies, at the level of $\sim$0.1-0.2 mag,
with respect to the other two SNe.
Since the bolometric fluxes for SN 1987A comprise two more IR bands, 
$L$ and $M$, we examined the possibility that these discrepancies could be 
due to this fact. For this, we excluded the $L$ and $M$ bands for 
SN 1987A and recomputed the bolometric flux in the same manner 
as for the other two SNe, i.e, by calculating $F_{IR}$ as the extrapolation
of a BB fit to $U-K$ photometry.
This exercise showed that, while
the BC corrections over the plateau phase do not change in any significant
way (lending support to the $F_{IR}$ derived from BB fits), by the end of 
the plateau and at later times the new BCs increase and
get closer to the other two SNe. The conclusion is that
the differences observed during the transition are due to an 
inaccurate estimate of $F_{IR}$ from the BB fits restricted to $U-K$ 
photometry. Adding $L$ and $M$ photometry at these late epochs do help
and provides a more accurate estimate of $F_{IR}$. Therefore, during the
transition we decided to exclude the BCs derived from SN 1999em and 2003hn.

A  good representation of the correlation between BC and colors can 
be obtained with polynomial fits of the form, 

\begin{equation}
BC(color)= \sum _{i=0}^{n} a_i \, (color) ^ i,
\end{equation}

\noindent where the order $n$ varies for each color. 
Table ~\ref{tbl-1} lists the coefficients obtained for the fit of each color, 
their range of validity and the number of  data points used. The fits 
have dispersions (rms) of 0.11 mag for  ($B-V$), 0.11 mag for ($V-I$) 
and 0.09 mag for ($B-I$) in the whole range (plateau plus transition 
to the radioactive tail).

The corresponding polynomials fits are also shown with  solid lines in 
Figure \ref{fig:BC1},  \ref{fig:BC2} and  \ref{fig:BC3}. As  a comparison, 
the dashed line in these Figures show the BC derived for a blackbody. 
The blackbody models represent well the data at early times (bluest colors),
but evidently differ from the atmosphere models and the observed
SNe at later epochs.

As argued in section ~\ref{sec:Lbol}, we have good reasons to trust more 
the atmopshere models than the early data of SN 1999em, so we decided 
to exclude the latter from our fits. Therefore, our calibration should
be considered more uncertain here. A further complication at early phases
is the steep dependence of the BC on color. This means that a slight 
error in the measurement of the color, such as that due to a poor 
extinction determination, could cause a significant error in the 
determination of the BC.
This problem is less pronounced if we use $(V-I)$  to estimate 
BC at these epochs.  We also tested if a bolometric correction with 
respect to the $R$ band instead of $V$ would 
improve the situation, but we did not find any improvement.

Using the coefficients given in Table ~\ref{tbl-1}, it is possible to 
derive a bolometric luminosity for any SN~II-P using only two (or three 
in the case of the $(B-I)$ color) optical filters. If one knows the 
extinction and the distance to the object the bolometric luminosity 
can be computed as,

\begin{equation}
\log L[ \mathrm {erg \,~s}^{-1}]=  - 0.4 \,\, [ BC(color) + V - A_{total} (V)  - 11.64  ]  + \log( 4 \, \pi D ^2 ),
\label{eq3}
\end{equation}

\noindent where $D$ is the distance in cm to the SN
and $A_{total}(V)$ is the total, 
host  plus Galactic, visual absorption. Note that combining this equation 
with equations (\ref{eq1}) and (\ref{eq2}), the luminosity becomes independent
of the arbitrary zero points chosen for the Vega magnitude scale, and it 
only depends on the observed flux density, color, extinction and distance.

\begin{figure}
\begin{center}
\includegraphics[angle=-90,scale=.40]{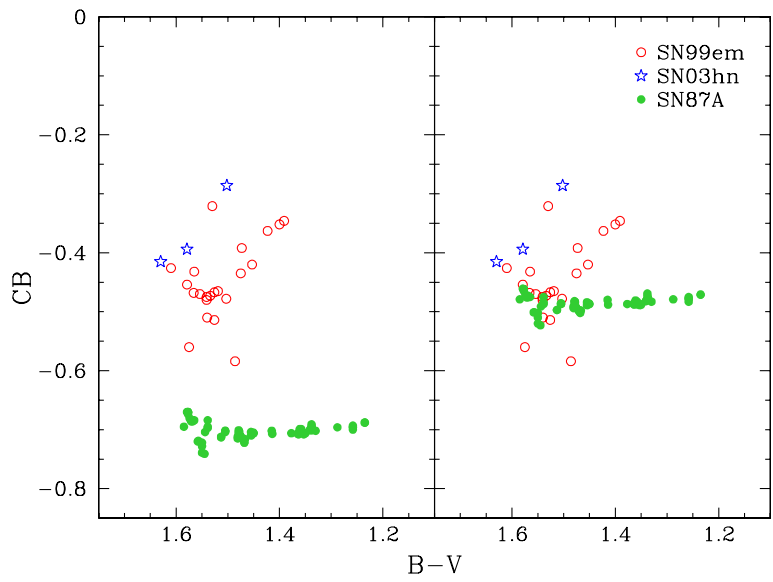}
\caption{({\em Left}) Bolometric corrections obtained during the
  radioactive tail phase versus $(B-V)$, as derived using
  $U$ thru $M$ photometry for SN 1987A (filled circles), and $U$ thru
  $K$ photometry for SN 1999em (open circles) and SN 2003hn (stars).
  ({\em Right}) Same as before, but excluding the $L$ and $M$ bands 
  for SN 1987A.  \label{fig:BCT}}
\end{center}
\end{figure}

The calibrations of BC versus colors shown above are only valid during 
the optically thick phases since they involve BB fits to the photometry.
In the nebular phase, we calculated BCs for the three SNe using the integrated
flux between the observed bands (i.e. $U-K$ for SN 1999em and SN 2003hn,
and $U-M$ for SN 1987A). We did not attempt to add any flux beyond these
limits since we did not have any physical model to extrapolate. As shown in
 the left panel of Figure \ref{fig:BCT} the BC for SN 1987A is almost
 independent of color,  
with a value of $-0.70$ mag and a scatter of only 0.015 mag.
The other two SNe yield BCs 0.2--0.3 mag higher, with a slight dependence 
on color. 
We investigated whether this differences could be due to the inclusion 
of the two additional bands for SN 1987A: we removed the $L$ 
and $M$ bands from the BC and, not surprisingly, the agreement proved
very good (Right panel of Figure \ref{fig:BCT}). We conclude that the
$L$ and $M$ contributions 
to the bolometric flux is not negligible at the nebular phase. Hence,
we take the value of $-0.70$ derived from SN 1987A as the best estimate 
of the BC at the onset of the nebular phase. 

We conclude this section with the claim that we have implemented
a robust method to estimate BCs for SNe~II-P which allows one
to derive bolometric luminosities. If we trust the late behavior of SN
1987A as being representative of SNe~II-P in general, our analysis
implies an overall accuracy of 0.05 dex in BCs. Clearly, it would be
interesting to check this result 
using L and M photometric data of other SNe~II-P, but such data are
currently unavailable. Our calibrations have the potential to be applied
to many SNe observed over a limited wavelength range.

\subsection{Effective Temperature-Color Relation}

Along with bolometric luminosity, 
the effective temperature is a critical parameter in the comparison of
observations with hydrodynamical models. 
Each model spectrum of E96 and D05 has an associated effective temperature
(T$_{eff}$), defined by the relation $L= 4 \pi  R_{ph}^2 \sigma T_{eff}^4$ 
where L is the input luminosity of the atmospheric models and
$R_{ph}$, the photospheric radius, is an output of the models. We
examined the dependence of 
T$_{eff}$ on $(B-V)$ and $(V-I)$ colors derived via synthetic
photometry from the model spectra, as described in section \ref{sec:Lbol}.
The purpose of this analysis was to provide a calibration between
temperature and color which could 
be used to easily derive estimates of T$_{eff}$ from observed colors
for any SN~II-P. Note, however, that T$_{eff}$ does
not have a direct physical meaning for this type of objects. It is
simply a convenient contact point between hydrodinamical models and
observations. 

\begin{figure}
\includegraphics[angle=-90,scale=.70 ]{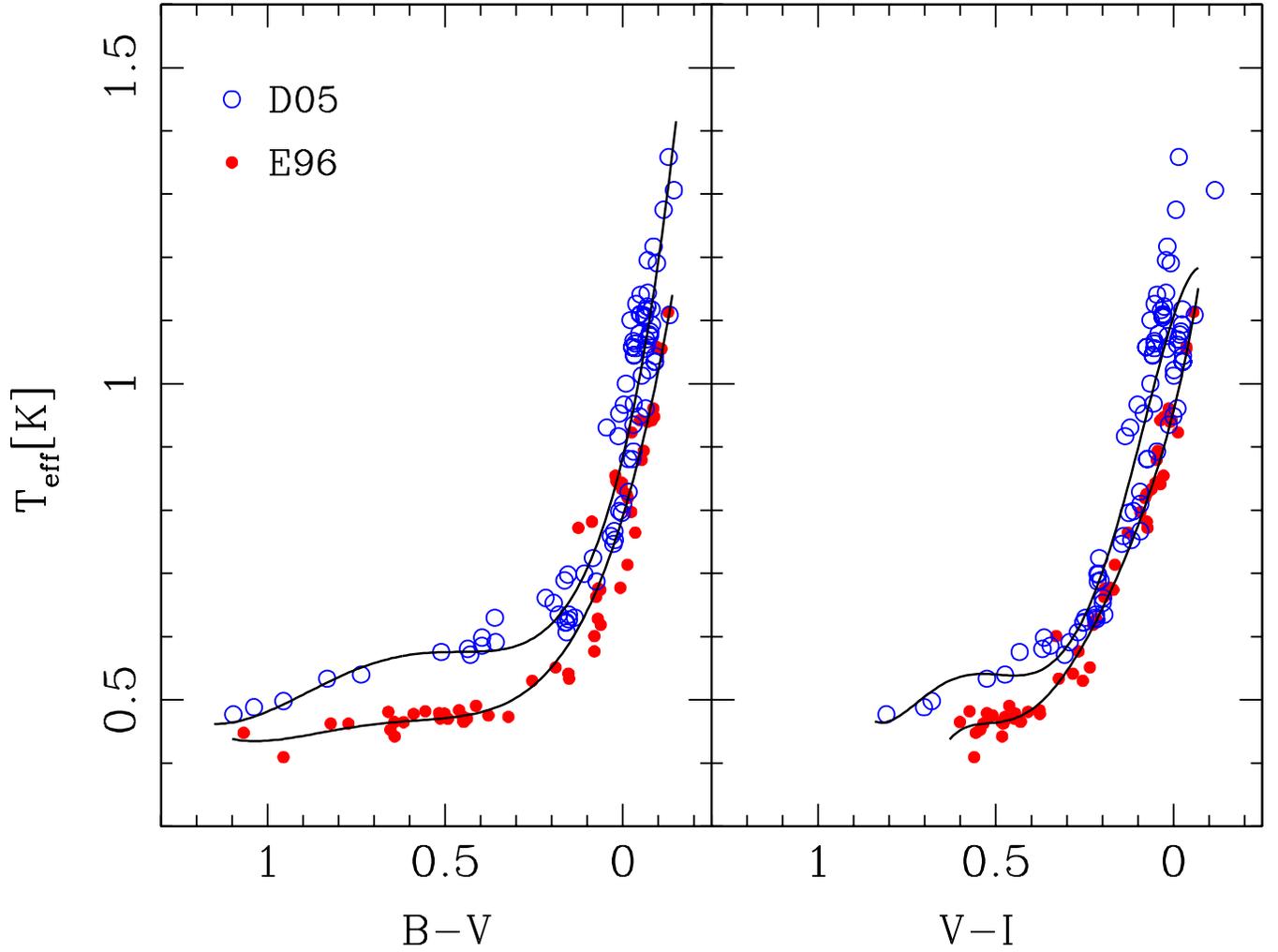}
\caption{ Effective  temperature versus $(B-V)$ ({\em left}) and
  $(V-I)$ ({\em right}) from the models of E96 
(filled circles) and D05 (open circles). The solid lines show polynomial fits 
for each model set. \label{fig:Teff}}
\end{figure}

Figure \ref{fig:Teff} shows the effective temperature versus synthetic 
$(B-V)$ and $(V-I)$ colors for E96 and D05 models. 
As expected, there is a tight correlation between these quantities 
for each set of models. At early epochs, when $(B-V) \lesssim 0.2 $ and
$(V-I) \lesssim 0.3$,
both models show consistent values of the effective temperature within
their internal dispersion. Later on, however, when the
plateau phase is well established, there are systematic differences in
the behavior of both sets, with the models of D05 giving   
larger effective temperatures. Similar differences have been reported 
in the literature with regard to the dilution factors calculated from 
both sets of models \citep{D05b,jones}, but there has been no clear
explanation for the discrepancies. Note that
 T$_{eff}$ ( $\propto \sqrt{R_{ph}}$) is an
output of the atmosphere model and depends on complicated 
details of the solution of radiation transport through the envelope,
such as non-LTE treatment of the different species and metal line
opacities.

In order to represent the correlation between  T$_{eff}$  and colors 
shown in Figure \ref{fig:Teff}, we fit polynomial functions 
of the form,

\begin{equation}
T_{eff}(color)[10^4 K]= \sum _{i=0}^{n} a_i \, (color) ^ i.
\end{equation}

\noindent The fits were done for each set of models separately and they 
are shown in Figure \ref{fig:Teff} with solid lines. The coefficients of the 
fits, ranges of validity for $(B-V)$ and $(V-I)$ colors, and dispersions 
are given in Table \ref{tbl-2}. The fits to the E96 models 
are characterized by a scatter of $\sim$500 K in $ (B-V)$  and $\sim$350 K
in $(V-I)$. For the D05 models the scatter is $\sim $670~K in $(B-V)$ 
and $\sim$800 K in $(V-I)$.  We do not  have any  strong argument to  
rule out either set of models. We therefore keep both results even if they 
show significant systematic differences.  But we notice that
the value of  T$_{eff}$ during the recombination phase (where it is nearly 
constant) appears to be somewhat underestimated  (T$_{eff} \sim 4600$ K) by 
E96. We recall that these calibrations are only valid until the end
of the plateau phase.

\begin{figure}
\epsscale{1.0}
\plotone{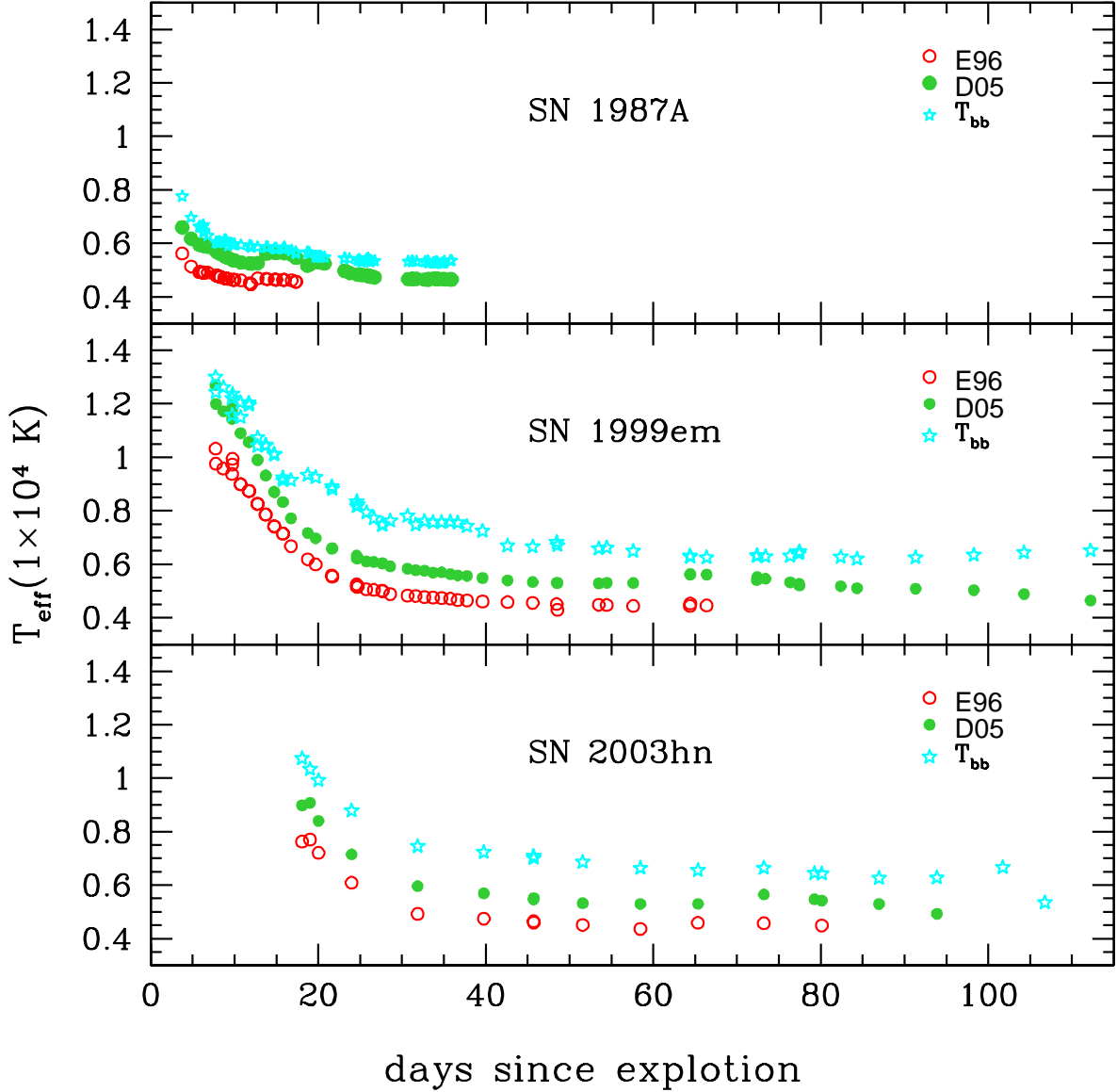}
\caption{Effective temperatures  for SN 1987A ({\bf upper panel}), 
SN 1999em ({\bf middle panel})  and SN 2003hn ({\bf bottom panel }) 
as a function of time since explosion
  calculated using the polynomial fits given in table \ref{tbl-2} for
  the models of E96 (open circles) and  D05 (filled circles). As a comparison
  we included in these plots the color temperatures obtained from the
  BB fits described in section \ref{sec:Lbol}. 
 \label{fig:Teff2}}
\end{figure}

In Figure \ref{fig:Teff2} we show how these calibrations work for estimating  
T$_{eff}$ for SN 1987A, SN 1999em, and SN 2003hn. 
As a comparison we included in these plots the color temperatures 
obtained from the BB fits described in section \ref{sec:Lbol}. Note that 
for the three SNe the color temperatures are greater than 
$T_{eff}$, which is expected for ``dilute'' atmospheres whose 
continuum opacity is dominated by electron scattering. As we said
before, the usefulness of these fits is that they allow one 
to obtain  T$_{eff}$ for any SN~II-P from their $(B-V)$ or $(V-I)$
colors, and thereby use this quantity  to compare with hydrodynamical
 models. Note that we could equivalently have chosen to calibrate
 R$_{ph}$ vs color. 

\section{Application to SNe~II-P Data} \label{sec:sample}

We  applied our calibrations for bolometric corrections and effective 
temperatures to a sample of 33 SNe~II-P with precise optical photometry. 
This sample of SNe~II-P was observed in the course of four systematic 
follow-up programs: the {\em Cerro Tololo Supernova Program} (1986-2003), 
the {\em Cal{\'a}n/Tololo Supernova Program} (CT; 1990-1993), 
the {\em Optical and Infrared Supernova Survey} (SOIRS; 1999-2000), 
and the {\em Carnegie Type II Supernova Program} (CATS; 2002-2003).  
Currently, all of the optical data have been reduced and they are in 
course of publication \citep{H09}. Additionally, we included four SNe 
from the literature: SN 1999gi \citep{2002AJ....124.2490L}, 
SN 2004dj \citep{2006MNRAS.369.1780V}, SN 2004et
\citep{2006MNRAS.372.1315S}, and SN 2005cs
\citep{Pa06,2006A&A...460..769T}. 

To calculate bolometric light curves for these SNe we only need  to 
apply equation (\ref{eq3}). Thus, we need to have (1) $B V I$ photometry, 
(2) extinction corrections due to our own Galaxy,  
(3) host-galaxy reddening corrections, and (4) distances.  

We obtained Galactic extinction from \citet{SFD98}.
Host-galaxy extinctions and distances for the present sample were 
calculated by \citet{F08} (see their Tables 2.2, 3.1, 4.4)
using the Standardized Candle Method. For two objects, SN 2003ef 
and SN 2005cs, \citet{F08} did not provide distances. For SN 2003ef 
we used the EPM distance estimated by \citet{jones} ,  
converted to the distance scale of \citet{F08} using 
the conversion coefficients  given by the authors. The resulting  
value was of  $48.55$ Mpc. For SN 2005cs we used the distance modulus 
of $\mu= 29.62$ mag  given by Pastorello et al 2006,
which corresponds to 8.4 Mpc.

\begin{figure}
\epsscale{1.0}
\plotone{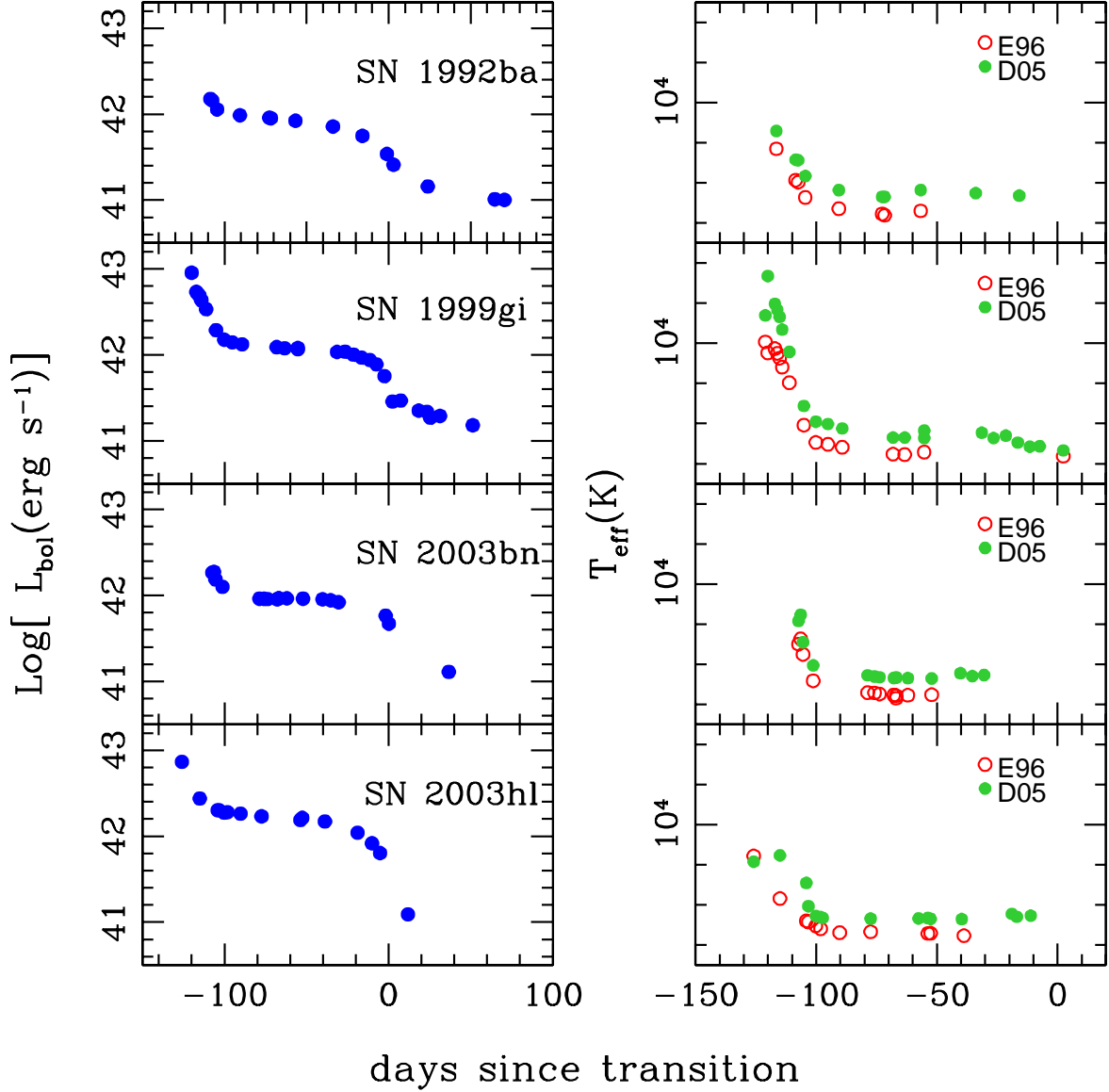}
\caption{(left) Bolometric luminosity curves for four well-observed
objects of our sample: SN 1992ba, SN 1999gi, SN 2003hn, and SN 2003hl.
(right) Effective temperatures for the same four SNe. The origin of time 
in these plots is the middle point between the plateau and the linear tail 
\citep{F08}.\label{fig:SNeLT}}
\end{figure}

We obtained bolometric light curves  and effective temperatures  
for the 33 SNe~II-P in the sample and  placed them in the same scale 
of time in order  to compare them. Given that we did not have an 
estimate of the explosion time for the   majority of the  SNe used 
in this analysis, we decided to use as origin of time the epoch defined 
by Olivares 2008 et al., i.e. the middle point of  the transition 
between the plateau and  the radioactive tail.  Figure \ref{fig:SNeLT}
shows the bolometric light curves and effective temperatures for a 
subsample of four objects. Bolometric light curves and effective 
temperatures for the remaining 29 SNe are provided only in the
electronic edition.

\begin{figure}
\epsscale{0.85}
\plotone{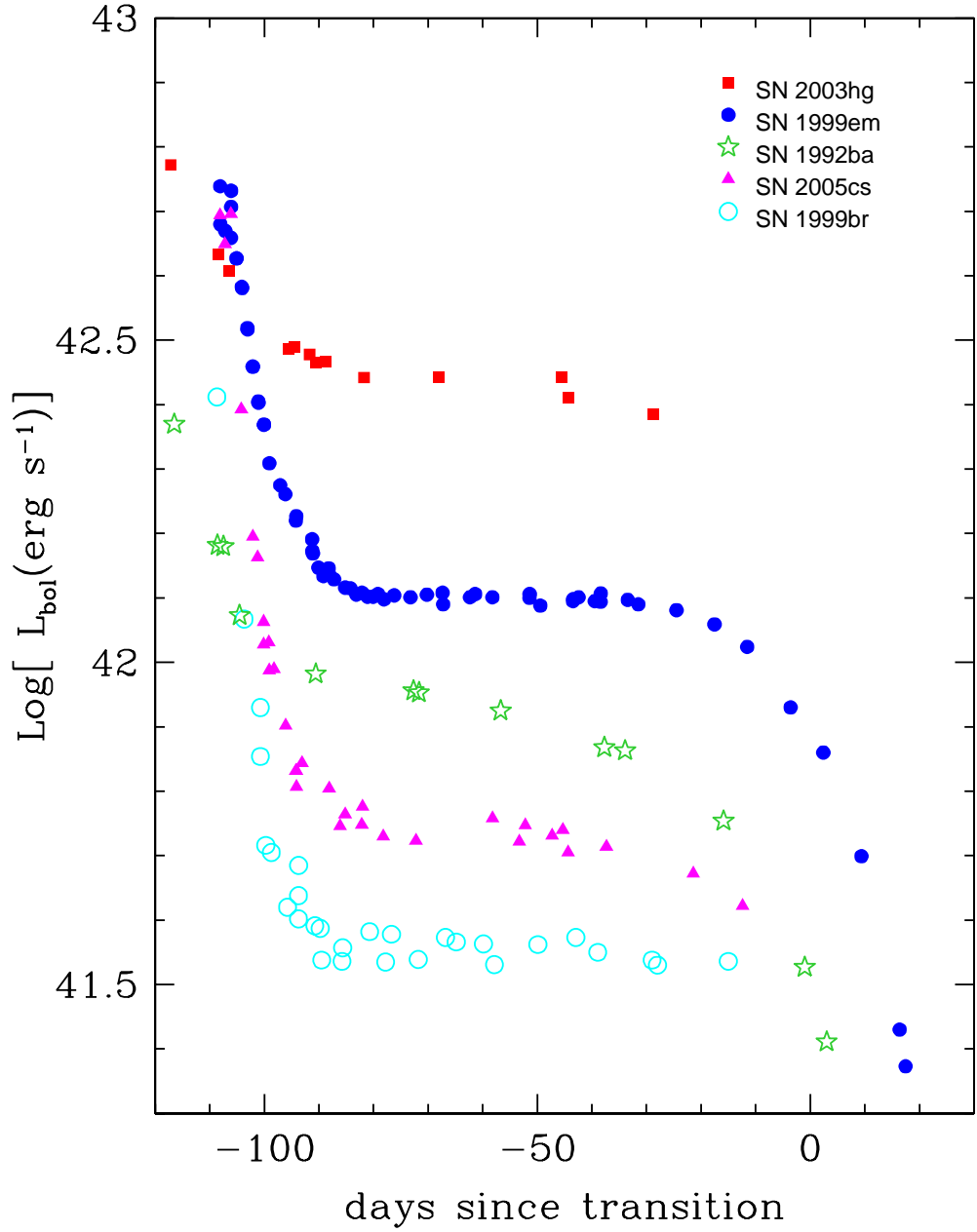}
\caption{Bolometric light curves for five SNe of our sample showing the range of variation of the plateau luminosities. \label{fig:SNePL}}
\end{figure}

A simple inspection of the resulting bolometric light curves of our 
sample reveals a high degree of heterogeneity
among these SNe II-P which can be also seen in the subsample Figure \ref{fig:SNeLT}.
 In Figure \ref{fig:SNePL} we compare five
exemplary SNe to illustrate the well known fact that SNe~II-P display 
a wide range of plateau luminosities \citep{H01}. The range of luminosities 
encompassed by our sample is $\sim$1 dex, equivalent to one order of magnitude 
of spread in their radiative energy output. The most frequent value of 
the plateau luminosity in our whole sample was $\sim 1.2\times10^{42}$
erg s$^{-1}$.

There are 16 SNe with data previous to the plateau phase, i.e. during 
the adiabatic cooling  phase. Among these, there are three subluminous 
objects, SN 1999br, SN 2003bl and SN 2005cs which, in comparison with
the rest, appear to have a steeper slope during the adiabatic cooling
phase and a flatter plateau (see Figure \ref{fig:SNePL}). In all cases
where we observed a plateau where the luminosity remained nearly constant,
or it slowly decreased, but it never increased. \citet{U07} has
predicted that part of the dense core of the massive star should be
ejected in order to produce a plateau phase as observed. On the
contrary, if the whole core remains in the compact remnant and the
density profile encountered by the shock wave in the envelope is
relatively flat, the resulting luminosity increases with time. Our
data suggest that all SN~II-P ejecta involve part of the dense core of
the progenitor star.

\begin{figure}
\begin{center}
\includegraphics[angle=-90,scale=0.6]{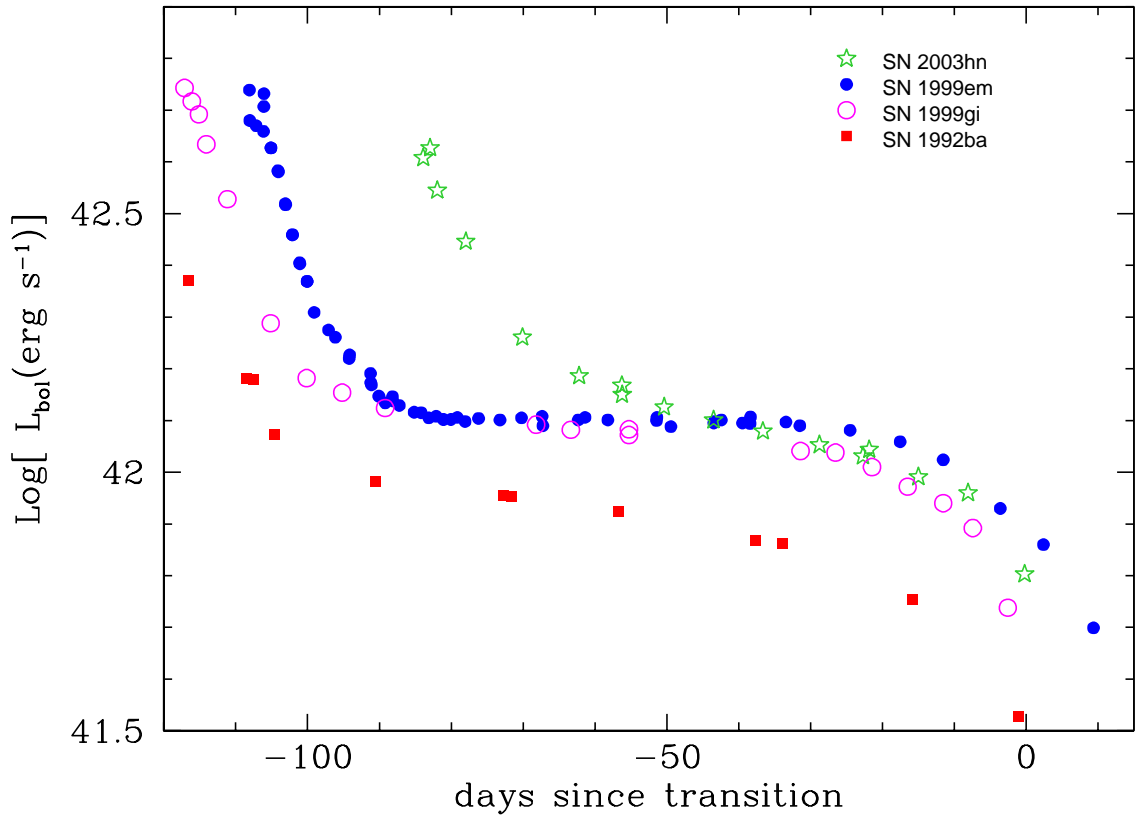}
\caption{Bolometric light curves for four SNe of our sample showing the range of variation of the plateau lengths. \label{fig:SNePD}}
\end{center}
\end{figure}

Twelve SNe with complete coverage between the adiabatic cooling phase and 
the transition to the radioactive tail showed a range of plateau  
durations between 75 and 120 days. In this study, we have defined the 
plateau phase as the part of the light curve on which the luminosity 
remains within 1 mag (or 0.4 dex) of the mean value calculated between 20 and 
80 days before the epoch in the middle point of the transition to the
radioactive tail \citep{F08}. Figure \ref{fig:SNePD} shows a comparison 
of the light curves of four SNe selected to illustrate the variety of
plateau lengths. It is known that the duration of the plateau is strongly 
related to the envelope mass of the progenitor object  more than 
other parameters such as explosion energy and radius
\citep{1983Ap&SS..89...89L,Po93}. See for example  the equation for
the plateau duration ($t_p$) given by \citet{Po93},$\; t_p \; \propto \;
R^{1/6} E ^{-1/6} M ^{1/2} $ where R is the initial radius of the progenitor
object, M is the ejected mass and E is the energy of the
explosion. Based on this, our results indicate that there is a variety
of envelope masses among the progenitors of
SNe~II-P. A quantitative assertion on this respect requires detailed
modeling of the data.

\begin{figure}
\epsscale{0.85}
\plotone{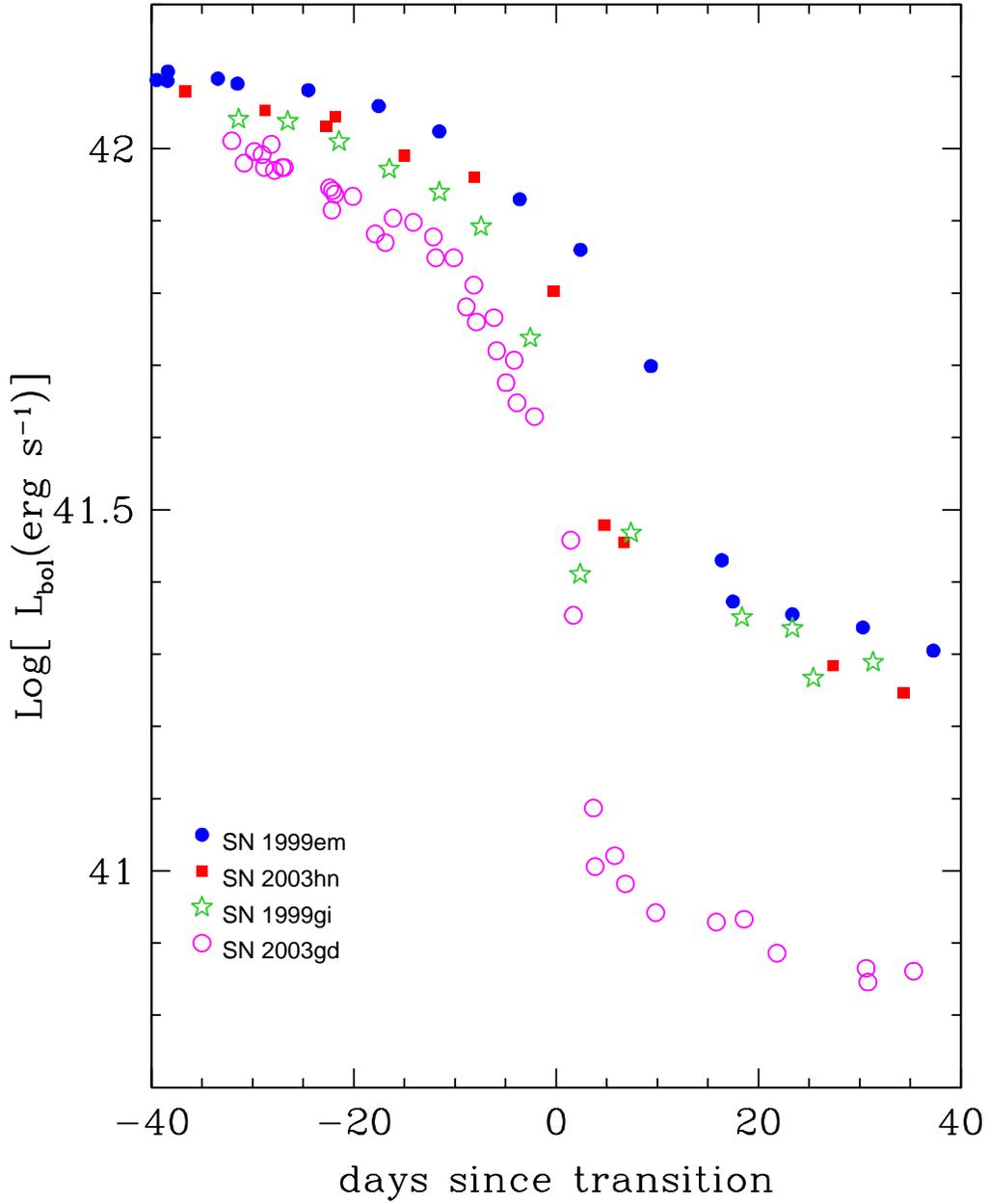}
\caption{Bolometric light curves of four SNe of our sample with different
transition properties. \label{fig:SNeT}}
\end{figure}

Finally, we compared the shape of the transition between the plateau 
phase and the radioactive tail. Unfortunately, there are only six SNe 
for which this transition is well sampled. In Figure \ref{fig:SNeT} we show 
this transition phase for four of the six SNe in this group. The 
luminosity drop is in the range of $\sim$ 0.8--1.12 dex. This behavior
is related to the $^{56}$Ni mass synthesized
in the explosion and its degree of mixing. The amount of  $^{56}$Ni 
sinthetized by the SN determines the height of the radioactive tail. 
More $^{56}$Ni mixing produces a more gradual transition between the plateau 
and the tail \citep{E94,U07}. The drop for SN 2003gd appears to be steeper  
and larger than for the other SNe, which is  indicative  of less mixing 
and a smaller $^{56}$Ni mass. From this 
preliminary study we can state that the $^{56}$Ni mass produced
and its mixing degree varies significantly among SNe~II-P.

We are currently working on the determination of physical parameters for 
the progenitor stars of this sample of SNe using a hydrodynamical code 
recently developed by us \citep{BBH09}. The goal of that work, to be 
published soon, is to analyze the  distribution of physical parameters 
for this sample of  SNe~II-P.

\section{Conclusions} \label{sec:conclusion}
Using data from three SNe with excellent observations  and two
atmospheric models we  derived   reliable calibrations for bolometric
corrections  and  effective temperature  from $B V I$  photometry
applicable to SNe~II-P. The characteristic scatter of the BC
calibration  during the plateau phase is of 0.11 mag which corresponds
to an uncertainty of  0.044 dex in bolometric luminosity. On the
radioactive tail we found that the BC was independent of color with a
value of $-0.70$ mag  and a scatter of 0.02 mag based only on the
behavior of  SN 1987A. We noticed the importance of including  $L$ and
$M$ photometry in order to derive an accurate BC during this phase. 

 We  emphasize that the largest uncertainties in our calibration of
 the  BC occur at the earliest epochs corresponding to   $(B-V)
 \lesssim 0.2 $, where the  discrepancies between models and data are
 the largest. During the rest of the  SN evolution, we found an
 overall very good  agreement between models and observations. A
 possible improvement to this calibration  could be achieved from
 early spectra of SNe~II-P in order to confirm the behavior shown by
 the models. In this sense  it is important to note the recent
 suggestion by  
\citet{Gal08} that the UV behavior of SNe~II-P is very uniform.
Another complication at early phases is the steep dependence of the BC
on color. We  point out that caution may be taken in the estimate of
the extinction in order to use the  BC to estimate luminosities at the
earliest epochs. In this sense, the BC of the  $(V-I)$ color  appears
to be less sensitive  to extinction and therefore offers a  better
calibration at early epochs. 

Regarding the effective temperature calibration, we found systematic
differences between the two sets of models considered but we did not
find any strong reason to rule out any of them. The characteristic
scatter of the T$_{eff}$ versus color calibration was $500$K for
$(B-V)$ and $350$K for $(V-I)$ in the case of the E96 models, while for 
the D05 models it was $670$K for $(B-V)$ and $ 800$K for $(V-I)$. 

 Based on these calibrations we derived bolometric luminosities and
 effective temperatures  for 33 SNe~II-P, which prove a useful
 resource to extract physical properties of this type of objects.  In
 this preliminary analysis we found that  the SNe~II-P  in our sample
 showed a range of $\sim$ 1 dex in plateau luminosity (Lp) with most
 of the SNe having Lp $\sim 1.2 \times 10^ {42}$ erg s$^{-1}$. Plateau
 durations ranged between 75 and 120 days, which indicates the
 existence of a variety of ejected masses among SNe~II-P.
 We also compared the shape of the transition
 between the plateau and the radioactive tail. We found that the size
 of the drop ranged between 0.8 and 1.12 dex indicating 
a variety of masses and degrees of mixing of $^{56}$Ni for this SN
sample. 

\acknowledgments
MB acknowledges support from  MECESUP UCH0118 program. MH obtained support from proyecto FONDECYT (grand 1060808), Millennium Center for Supernova Science through grant P06-045-F, Centro de Astrof\'\i sica FONDAP 15010003, and Center of Excellence in Astrophysics and Associated Technologies (PFB 06).

\clearpage

\clearpage

\clearpage

\clearpage

\clearpage
\begin{deluxetable}{ccrrrrrrrrcrl}
\tablecaption{Coefficients of the fits to BC(color)\tablenotemark{a}. \label{tbl-1}}
\tablewidth{0pt}
\tablehead{
\colhead{a$_i$} & \colhead{$B-V$} & \colhead{$V-I$} & \colhead{$B-I$} }  
\startdata
a$_0$ &  -0.823 &  -1.355  & -1.096  \\
a$_1$ &   5.027 &   6.262  &  3.038  \\
a$_2$ & -13.409 &  -2.676  & -2.246  \\
a$_3$ &  20.133 & -22.973  & -0.497  \\
a$_4$ & -18.096 &  35.542  &  0.7078 \\
a$_5$ &   9.084 & -15.340  &  0.576  \\
a$_6$ &  -1.950 &  $ \cdots $  & -0.713  \\
a$_7$ &   $ \cdots $ &   $ \cdots $  &  0.239  \\
a$_8$ &   $ \cdots $ &   $ \cdots $  & -0.027  \\
\hline
ranges  & $[-0.2 ,1.65] $  & $ [-0.1, 1] $ & $ [-0.4, 3] $ \\
No. points &  512 &  465 & 512  \\
rms [mag]    &  0.113      &  0.109        &   0.091  \\

\enddata
\tablenotetext{a}{$BC(color)= \sum _{i=0}^{n} a_i \, (color) ^ i$}
\
\end{deluxetable}

\begin{deluxetable}{ccrrrrrrrrcrl}
\tablecaption{Coefficients of the fits to T$_{eff}$(color)\tablenotemark{a}. \label{tbl-2}}
\tablewidth{0pt}
\tablehead{
\colhead{} & \colhead{$B-V$} &\colhead{$B-V$}& \colhead{$V-I$}& \colhead{$V-I$} \\
\colhead{a$_i$} & \colhead{$E96 $ } & \colhead{$D05 $ } & \colhead{ $ E96 $}& \colhead{ $ D05 $} }
\startdata
a$_0$ &  0.790 &  0.884  &  0.957  &  1.106   \\
a$_1$ & -1.856 & -2.340  & -2.254  & -1.736   \\
a$_2$ &  4.055 &  6.628  &  5.922  & -6.403   \\
a$_3$ & -3.922 & -8.456  & -18.476 &  33.762  \\
a$_4$ &  1.368 &  4.619 &  36.058 & -48.260  \\
a$_5$ &  $\cdots$ & -0.849 & -25.291 &  22.362  \\

\hline
ranges    & $[-0.2,1.15] $  & $[-0.2,1.15]$ & $[-0.1,0.65]$ & $[-0.07,0.83]$  \\
rms [K] &  500 & 670 & 350 &  800 \\

\enddata
\
\tablenotetext{a}{$T_{eff}(color)[10^4 K]= \sum _{i=0}^{n} a_i \, (color) ^ i$}
\
\end{deluxetable}

\end{document}